\documentclass[pre, preprint]{revtex4}
\usepackage{graphicx}
\usepackage{epsfig}
\usepackage{rotating}

\begin{document}

\title{ On the Helix-coil Transition in Alanine-based Polypeptides in Gas Phase}

\author{Yanjie Wei} 
\email{yawei@mtu.edu}
\affiliation{ Department of Physics, Michigan Technological University, 
             Houghton, MI 49931, USA}

\author{Walter Nadler}
\email{wnadler@mtu.edu}
\affiliation{Department of Physics, Michigan Technological University, 
             Houghton, MI 49931, USA}

\author{Ulrich H.E. Hansmann} 
\email{hansmann@mtu.edu, u.hansmann@fz-juelich.de}
\affiliation{Department of Physics, Michigan Technological University, 
             Houghton, MI 49931, USA}
\affiliation{John-von-Neumann Institute for Computing, 
             Forschungszentrum J\"ulich, D-52425 J\"ulich, Germany}

\date{\today}

\begin{abstract}
Using multicanonical  simulations, we study the  effect  of charged end groups on helix formation in 
alanine based polypeptides.  We confirm earlier reports that neutral poly-alanine   
exhibits a pronounced  helix-coil transitions  in gas phase simulations.  Introducing
a charged Lys$^+$  at the C-terminal  stabilizes the $\alpha$ helix and leads to a higher transition temperature. On the other hand, adding the Lys$^+$ at the N-terminal inhibits helix formation.
Instead, a more globular structure was found. These results are in agreement with recent experiments on alanine based polypeptides in gas phase. They indicate that present
force fields describe accurately the intramolecular interactions in proteins.

\end{abstract}


\maketitle

\section{Introduction}
First principle calculations of protein folding have remained 
a computationally hard problem. This is in part because 
the energetic and entropic balance of a solvated protein in equilibrium
is complicated to express on a computer. Other problems arise from the 
inherent difficulties in sampling the rough energy landscape of a protein.
It is not a priori clear what contributes most to the numerical
shortcomings:   poor sampling techniques, lack of accuracy in
the  force fields that describe the intramolecular forces in a protein, 
or the crude approximations in the modeling of protein-solvent interaction. 
The introduction of high-resolution ion mobility measurements, which allow experiments to be performed on biological molecules in gas phase
 \cite{Jarrold_Dugourd,Jarrold_Hudgins},  has opened a way to simplify the problem.
By comparing simulations with gas-phase experiments one can avoid the
complications arising from the modeling of protein-solvent interactions. Hence, gas-phase
experiments offer a way to test whether current force fields describe correctly
the intramolecular forces and the ability of  present simulation techniques 
to sample accurately low-energy configurations.

Such evaluation of methods and energy functions through comparison with gas phase
experiments is done best with simple systems.  Often used  are homopolymers of
amino acids. For instance, the helix-coil transition has been studied extensively with
poly-alanine ~\cite{Jarrold_ala, Peng2003, Jarrold_ala2}. Numerical results  indicate a strong propensity to form helices if the temperature is below a certain critical value. On the other hand, the experimental observations seemed to contradict those theoretical results. For this reason, it was
conjectured in Ref.~\onlinecite{Jarrold_Kinnear}  that the strong  helix-coil transition 
for polyalanine in gas-phase simulations is an artifact of the utilized energy function.  
However, the deviations may also  result from differences in the systems studied. 
Gas phase experiments require partially charged molecules while in  simulations 
it is commonly assumed that the molecule has no total charge. 

In order to settle this question and quantify the effects of charged end groups we have 
simulated three alanine based polypeptides in gas phase. The first one is 
Ala$_{10}$ which we are familiar with from earlier  
work ~\cite{Peng2003, Peng2002, Alves, Hansmann1999}.  Simulations of this 
neutral molecule are supplemented by those of two charged polypeptides.  
In  Ace-Lys$^+$-Ala$_{10}$  the charged group sits at
 the N terminal while in  Ace-Ala$_{10}$-Lys$^+$  the charge sits at the C-terminal.
 Our results confirm the experimental findings that capping the N terminal by a 
 positively charged Lysine destabilizes the 
helix and leads to  a  more globular low-temperature structure, 
while a C-terminal Lys$^+$  can stabilize the $\alpha$ helix.

\section{Methods}
Our simulations utilize  the ECEPP/3 force field \cite{EC} as
implemented in the 2005 version of the program package SMMP \cite{SMMP,SMMP05}.
 Here the interactions between the atoms within the homopolymer chain are approximated
 by a sum $E_{ECEPP/3}$ consisting of electrostatic energy $E_C$, a  Lennard-Jones term $E_{LJ}$,
 hydrogen-bonding term $E_{HB}$ and a torsion energy $E_{Tor}$:
 \begin{eqnarray}
  E_{\text{ECEPP/3}} &=& E_C + E_{LJ}  + E_{HB} + E_{Tor} \nonumber \\
  &=&  \sum_{(i,j)} \frac{332 q_i q_j}{\epsilon r_{ij}} \nonumber \\
 &&   + \sum_{(i,j)} \left( \frac{A_{ij}}{r_{ij}^{12}} - \frac{B_{ij}}{r_{ij}^6} \right) \nonumber \\
 &&   + \sum_{(i,j)} \left( \frac{C_{ij}}{r_{ij}^{12}} - \frac{D_{ij}}{r_{ij}^{10}} \right) \nonumber \\
  && + \sum_l U_l ( 1\pm \cos(n_l \xi_l)) \;,
\label{energy}
\end{eqnarray}
where $r_{ij}$ is the distance between the atoms $i$ and $j$,  $\xi_l$ is the $l$-th torsion
angle, and energies are measured in Kcal/mol.

The above defined energy function leads  to a landscape that is characterized by a multitude of minima separated by high barriers.  In order to enhance sampling we therefore utilize  the multicanonical approach \cite{MU,MU2} as described in Ref.~\onlinecite{HO}.  Configurations are weighted 
with a term $w_{MU} (E)$ determined iteratively\cite{Berg96} such
that the probability distribution obeys
\begin{equation}    P_{MU}(E) \propto n(E) w_{MU}(E) \approx const~,
\end{equation}
where $n(E)$ is the spectral density of the system. Thermodynamic averages of an 
observable $<O>$ at temperature $T$ are obtained by re-weighting \cite{FS}:
\begin{equation} 
<O>(T) = \frac{\int dx \; O(x) e^{-E(x)/k_BT} / w_{MU}[E(x)]}
         {\int dx \; e^{-E(x)/k_BT} / w_{MU}[E(x)]}
\end{equation}
where $x$ counts the configurations of the system.

Various quantities are measured during simulations for further analysis. 
These include the energy $E$,
the radius of gyration $R_{gy}$ and the end-to-end distance $R_{ee}$ as measures of the geometrical size, and the helicity.
Here we define the helicity by the number of backbone hydrogen bonds, $n_{b-HB}$,
between residues $(i,i+4)$, which is characteristic for $\alpha$ helices.
We also monitored the number of non-backbone hydrogen bonds, $n_{nb-HB}$, as a measure of tertiary interactions. Finally we recorded  the lowest energy configurations.

\section{Results and Discussions}
As we are interested in the helix-coil transition in alanine-based polypeptides, 
we start our investigation with an analysis of the average helicity for the three 
peptides, measured by the number of backbone hydrogen bonds $\left<n_{b-HB}\right>$. 
This quantity is displayed as a function of temperature in Fig.~\ref{Helicity}. 
For the neutral molecule, Ala$_{10}$,  a sharp transition is observed that separates 
a disordered high temperature phase with vanishing helicity from an ordered low-temperature
phase where most configurations have high helicity. Not surprisingly,
the minimal energy configuration exhibits a fully formed $\alpha$ helix and
is displayed in Fig.~\ref{lowConfig}a.  
Displaying the specific heat  as a function of temperature in Fig.~\ref{speHeat}, 
we observe a  sharp peak  which indicates that this helix-coil
transition takes place at a ``critical" temperature   $T_C$ = 462K $\pm$ 5K .
Note that this value is about 40 K higher than the corresponding temperature  
427K $\pm$ 7K for the older ECEPP/2 force field \cite{Hansmann1999} 
although the two transitions differ little otherwise  (data not shown).

In order to study the  effect  of charges on the stability of an 
$\alpha$ helix, we perform a second simulation with  Ace-Ala$_{10}$-Lys$^+$ .
In this molecule, a positively charged Lys residue is added  at the C-terminal end of the chain.
The measured values for the helicity of  Ace-Ala$_{10}$-Lys$^+$  are similar to Ala$_{10}$, 
forming again a single extended helix  upon lowering the temperature (see Fig.~\ref{Helicity}). 
Note that the peak in specific heat 
(Fig.~\ref{speHeat}) is with  $T_C$= 557 $\pm$ 19K at an $\approx 100$ K higher temperature, and is much wider.
  This increase in the transition temperature indicates that the  charged Lys$^+$ at the 
  C-terminal favors the $\alpha$ helix through stabilizing the helix dipole. Parts 
  of the $\alpha$ helix survive even up to higher temperatures widening 
  the transition regime considerably.
  The stable $\alpha$ helix structure is also seen in the minimal energy configuration  displayed in
Fig.~\ref{lowConfig}b. Note  the additional hydrogen bond formed 
by the charged end group.  Our results are consistent with recent experiments that
also indicate stabilization by charged groups capping the C terminus and 
by the interaction of the charge with the helix dipole\cite{stability}. 

The third peptide  that we have studied is   Ace-Lys$^+$-Ala$_{10}$.  
Here the charged Lys is added at the N-terminal. For this peptide one finds only  
a small peak in the  specific heat  (see Fig.~\ref{speHeat}), 
 just below that of Ala$_{10}$ ($T_C$= 449 $\pm$ 37K),
  and a shoulder at higher temperatures. At low temperatures,  
its helicity (displayed in Fig.~\ref{Helicity})   is substantially lower
 than for the previous two peptides. Fig.~\ref{lowConfig}c displays the 
minimal energy configuration. Only a small partial helix is seen in the central region, 
while the end groups coil and  form a more compact globular structure.
The inlay in Fig.~\ref{Helicity} indicates that non-helical hydrogen bonds,  stabilizing 
 the globular structure, are formed already around 600K, above the small peak in spec. heat.
This formation of the globular structure as well as the joining up of the end groups 
is corroborated by the collapse of the molecule in that temperature range, 
as seen in the radius of gyration, $\left<R_{gy}\right>$, as well as in the 
end-to-end distance,  $\left<R_{ee}\right>$, in Fig.~\ref{rgy}. For comparison, 
in the uncharged molecule the drop in $\left<R_{gy}\right>$ occurs at lower 
temperatures, i.e. at the helix-coil transition, with practically no change in the 
end-to-end distance (Ace-Ala$_{10}$-Lys$^+$, the charged helix-forming variant, 
shows a similar behavior). In contrast, the formation of the helical sub-segment 
in Ace-Lys$^+$-Ala$_{10}$ does not show up significantly in the geometrical 
measures of that molecule. 
Hence, our data indicate that there is  no well-defined helix-coil phase transition for 
Ace-Lys$^+$-Ala$_{10}$. Instead, the shoulder in the specific heat rather 
marks the collapse of the molecule, while the peak at lower temperatures 
marks the formation of the helical sub-segment.
The reason for the suppressed and only partial helix-formation is the unfavorable 
interaction between the  N-terminal Lys$^+$ and the helix dipole, destabilizing the extended helix.
This conjecture is again supported by experimental finding in gas phase where the relative cross sections of  this peptide also indicate configurations that are more compact than helices\cite{Jarrold_Kinnear}.

\section{Conclusion}

This study has been motivated by a discrepancy between gas-phase experiments and first principle calculations of poly-alanine that either could indicate shortcomings of our energy function, the ECEPP
force field, or differences between the systems studied. 
As the original work in Refs.~\onlinecite{Hansmann1999} focused on
neutral molecules, while  the experiments studied charged molecules\cite{Jarrold_Kinnear}, we have now compared
simulations of polyalanine with either no charge, the charge at the C-terminus, or at the N-terminus.
Our simulation for the charged molecules are now in  agreement with the experimental results.
Our results underline two points. First, when comparing gas phase experiments with simulations
one has to make sure that the experimental settings are adequately described in the computer
experiment. Secondly, the agreement between gas phase experiment and simulation
suggests that the current generation of force fields describes the intramolecular forces within a protein sufficiently accurate to allow for correct secondary structure formation. The often observed failure of protein simulations in finding the correct structure
(see, e.g. Ref.~\onlinecite{PTH}) therefore  result likely from insufficient sampling and poor representations
of the protein-solvent interaction.

{\em Acknowledgments }
Support by a research grant (CHE-0313618) of the National Science Foundation (USA) is acknowledged.


%

%
%
\clearpage
{\huge Figure captions:}

\begin{description}
\item{Fig.~1:} Average helicity, measured by the number of backbone hydrogen bonds, $\left<n_{b-HB}\right>$, as function of temperature $T$ for the three polypeptides, obtained from a multicanonical simulation
with $5\times 10^6$ sweeps.
(open circles) Ala$_{10}$, 
(open squares) Ace-Ala$_{10}$-Lys$^+$, and 
(filled squares) Ace-Lys$^+$-Ala$_{10}$;
error bars are included and are mostly about the symbol size or less.
The inlay shows the non-backbone hydrogen bonds $\left<n_{nb-HB}\right>$, a measure of tertiary interactions.

\item{Fig.~2:} Specific heat $C(T)$ as function of temperature $T$ for the three polypeptides, obtained from a multicanonical simulation with $5\times 10^6$ sweeps. Symbols as in Fig. 1. Error bars are included and are mostly about the symbol size or less.

\item{Fig.~3:} Average radius of gyration $\left<R_{gy}\right>$ as function of temperature $T$ for the three polypeptides, obtained from a multicanonical simulation with $5\times 10^6$ sweeps. Symbols as in Fig. 1.
Error bars are included and are mostly about the symbol size or less.
The inlay shows the average end-to-end distance, $\left<R_{ee}\right>$.

\item{Fig.~4:} Minimal energy configurations of (a) Ala$_{10}$, 
(b) Ace-Ala$_{10}$-Lys$^+$, and (c) Ace-Lys$^+$-Ala$_{10}$,
obtained by minimizing the lowest energy configuration found during the simulation. The dashed lines show hydrogen bonds. The pictures have been drawn with VMD\cite{vmd}.

\end{description}
%


%
\setcounter{figure}{0}
%
\clearpage

\begin{sidewaysfigure}
    \includegraphics[width=1.0\columnwidth]{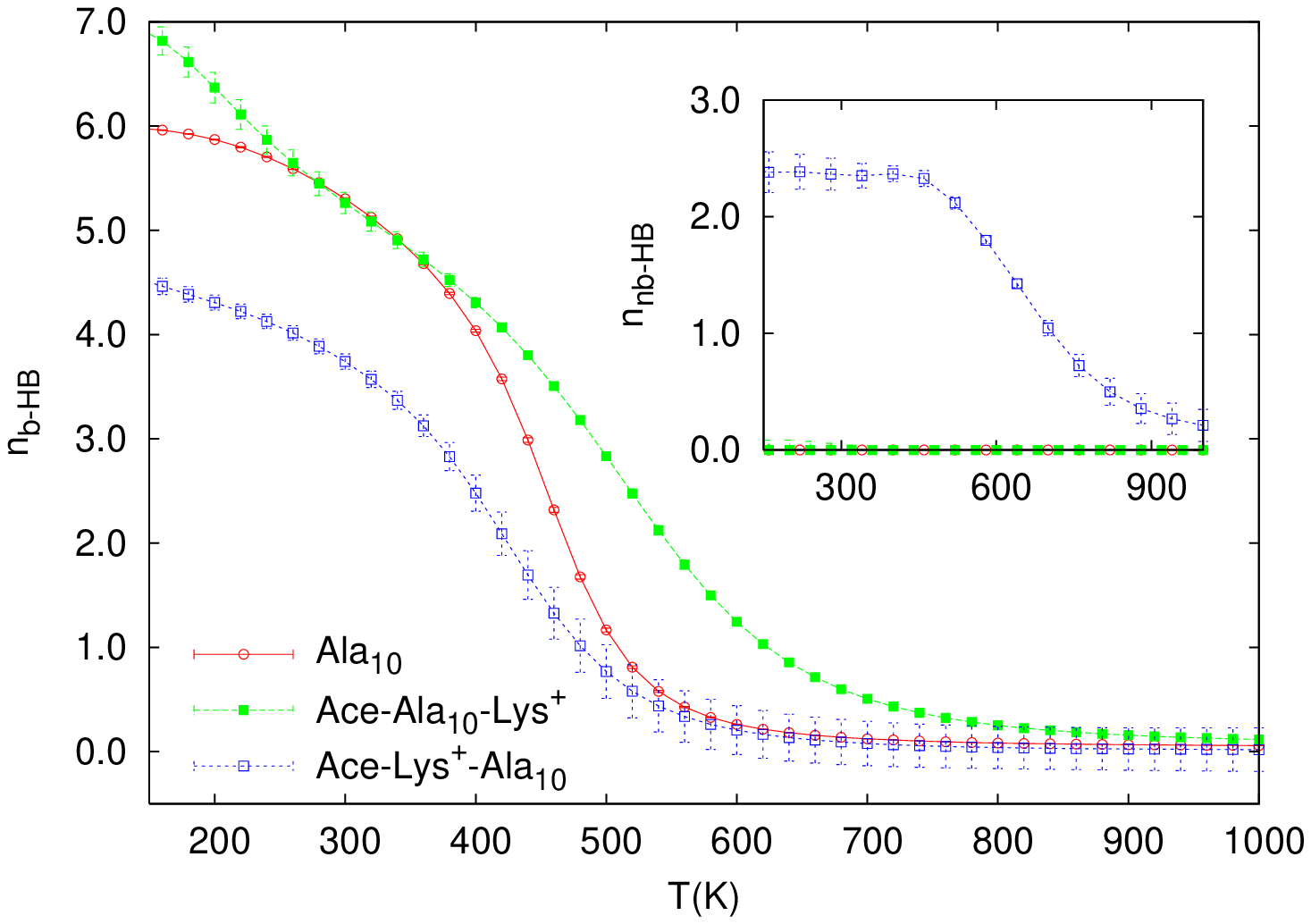}
\caption{\label{Helicity}}
\end{sidewaysfigure}

\begin{sidewaysfigure}
    \includegraphics[width=1.0\columnwidth]{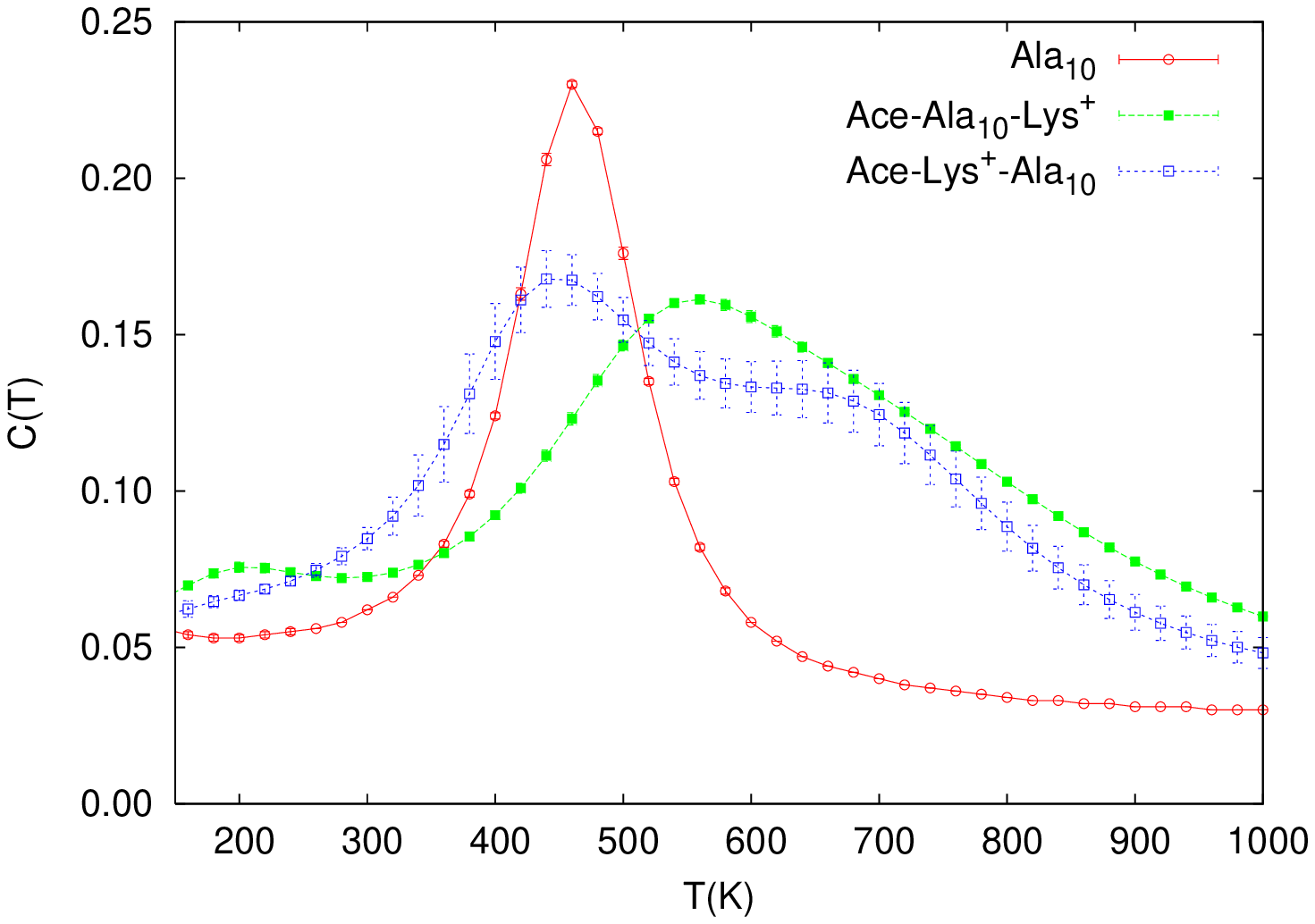}
\caption{\label{speHeat}}
\end{sidewaysfigure}

\begin{sidewaysfigure}
    \includegraphics[width=1.0\columnwidth]{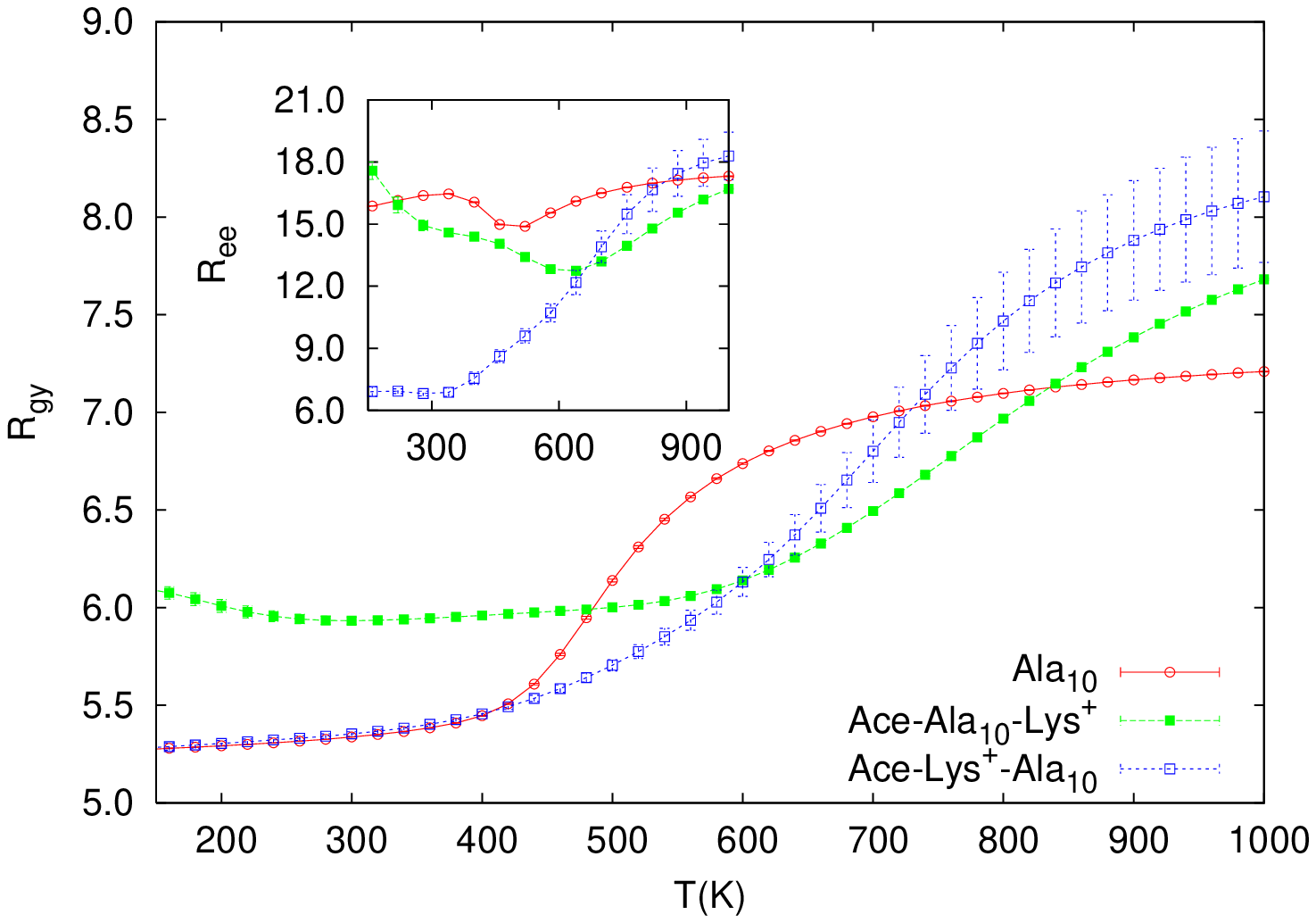}
\caption{\label{rgy}}
\end{sidewaysfigure}

\begin{figure}
    \includegraphics[width=1.0\columnwidth]{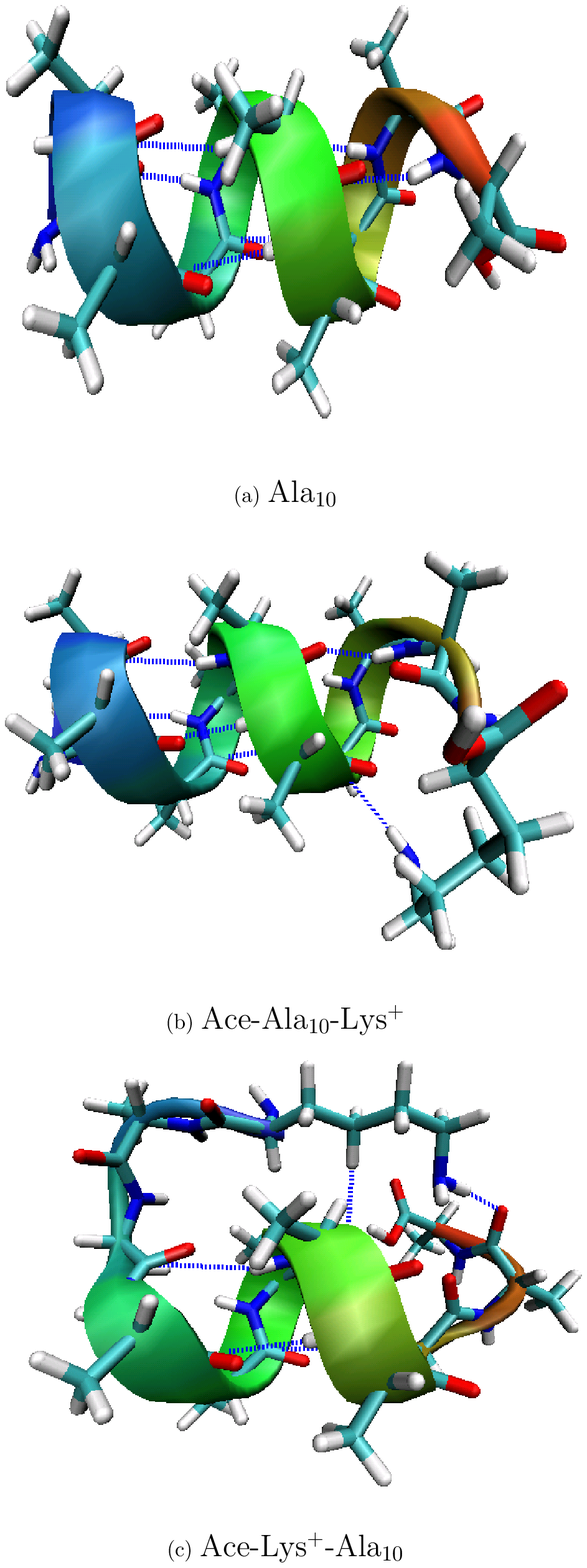}
\caption{\label{lowConfig}}
\end{figure}

\end{document}